\documentclass{ws-ijmpd}

\usepackage{graphicx}

\usepackage{amssymb}

\begin{document}

\markboth{Pereyra,P.H.}{The Schwarzschild Static Cosmological Model}

%
\catchline{}{}{}{}{} %

\title{THE SCHWARZSCHILD STATIC COSMOLOGICAL MODEL}

\author{PEREYRA,P.H.} 

\address{UNISC, Av.Independ\^{e}ncia 2293,CEP 96815-900,Santa.Cruz do Sul,RS,Brasil\\
FACOS, Rua 24 de Maio,141,Centro,95520-000,Os\'{o}rio,RS\\
FACCAT, Av.Oscar Martins Rangel,4500,(RS115),95600-000,Taquara,RS\\
pereyraph@gmail.com}

\maketitle

\begin{abstract}
The present work describes an immersion in 5D of the interior Schwarzschild solution of the general relativity equations. The model-theory is defined in the context of a flat 5D space-time-matter Minkowski model, using a Tolman like technique, which shows via Lorentz transformations that the solution is compatible with homogeneity and isotropy, thus obeying the cosmological principle. These properties permit to consider the solution in terms of a cosmological model. In this model the universe may be treated as an idealized star with constant density and variable pressure, where each observer can be ``center" of the same. The observed red-shift appears as a static gravitational effect which obeys the sufficiently verified and generally accepted square distance law. The Buchdahl stability theorem establishes a limit of distance observation with density dependence.
\end{abstract}


\begin{history}
\received{2 January 2006}
\comby{D. Hadjimichef}
\end{history}

\section{Introduction}

Recent results\cite{ABO,LAC,MCM,WES} in cosmology show the possibility of an immersion of cosmological models (static and non-static) in flat 5D spaces (Riemann flat), opening doorways for a better understanding of the behavior of such models in a geometrical sense \cite{LAC} as well as indicating new perspectives in the quantization issue \cite{BER,TRI} and unifying them with one geometric origin. However, already in 1930 Tolman \cite{TOL1,TOL2}, one of the pioneers in the field, applied immersion techniques in 5D to show the homogeneity and the geometry of static and non-static cosmological models.
According to the procedures of Tolman \cite{TOL3} and others \cite{FRI,ROB}, the only possible static models are reduced to three, Einstein, De Sitter and Special Relativity, that satisfy the cosmological principle (homogeneity and isotropy). In this work, using a similar technique used by Tolman, the possibility of another alternative for a static cosmological model is shown not contemplated by the previous three options, given by the Schwarzschild interior solution \cite{SCH} for the General Relativity theory (an idealized model for a star as incompressible fluid, with spherical symmetric distribution, of radius $r_1$, constant density $\rho _{00}$) and with metric in natural units (G=1, c=1) given by
\begin{equation}\label{eq1}
ds^2=-\frac{dr{}^2}{1-\frac{r\,{}^2}{R^2}}-r\,{}^2d\theta
\,{}^2-r\,{}^2\sin^2\theta \,\,d\phi \,{}^2+\left( {A-B\sqrt{1-\frac{r\,{}^2}{R^2}}}\right) ^2dt\,^2\;,  
\end{equation}
where
\begin{equation}\label{eq2}
R^2=\frac 3{8\pi \rho _{00}},\,A=\frac 32\sqrt{1-\frac{r_1^2}{R^2}}%
,\,\,B=\frac 12,\,\,m=\frac{4\pi }3\rho _{00}r_1^3  
\end{equation}
which defines the invariant line element.

\section{Transformations}

\subsection{Minkowski Metric 5D}

In order to accomplish an immersion of the solution (\ref{eq1}) in a Minkowski 5D (M-5) metric, we proceed in two steps. Initially, the known transformations, already used by \cite{TOL2}, are given as follows,
\begin{equation}\label{eq3}
r=R\,\sin (\chi )
\end{equation}
\begin{equation}\label{eq4}
\alpha =R\sin \left( \chi \right) \sin \left( \theta \right) \cos \left(
\phi \right)
\end{equation}
\begin{equation}\label{eq5}
\beta =R\sin \left( \chi \right) \sin \left( \theta \right) \sin \left( \phi
\right)
\end{equation}
\begin{equation}\label{eq6}
\gamma =R\sin \left( \chi \right) \cos \left( \theta \right)  
\end{equation}
\begin{equation}\label{eq7}
\zeta =R\cos (\chi )
\end{equation}
that take the spatial part of (\ref{eq1}) as a Euclidian metric 4D,
\begin{equation}\label{eq8}
ds^2=-d\alpha ^2-d\beta ^2-d\gamma ^2-d\zeta ^2+\left( {A-B}\frac{{\zeta }}
R\right) ^2dt{}^2. \; . 
\end{equation}
In the following, the system of partial differential equations are solved,
\begin{equation}\label{eq9}
-d\delta ^2+d\varepsilon ^2=-d\zeta ^2+\left( {A-B}\frac{{\zeta }}R\right)
^2dt{}^2.  
\end{equation}
for the functions $\delta \left( \zeta ,t\right) $ and $\varepsilon \left(
\zeta ,t\right) $, yielding as result the transformations
\begin{equation}\label{eq10}
\delta =R\cosh \left( \frac tR\right) \left( A-B\frac{{\zeta }}R\right)
\end{equation}
\begin{equation}\label{eq11}
\varepsilon =R\sinh \left( \frac{{t}}R\right) \left( {A-B}\frac{{\zeta }}%
R\right) ,
\end{equation}
and therefore (\ref{eq8}) changes into
\begin{equation}
ds^2=-d\alpha ^2-d\beta ^2-d\gamma ^2-d\delta ^2+d\varepsilon ^2,
\label{eq12}
\end{equation}
which assumes the form of the Minkoweski 5D metric (Riemann Flat), as desired.

\subsection{Lorentz 5D Tranformations}

In the following we show that for a group of transformations in 5D space, the line element (\ref{eq1}) is homogeneous, in other words, invariant under translations and by virtue isotropic and non-homogeneous due to a variable pressure \cite{TOL2}). The apparent non-homogeneity is due to a fictitious effect of the 4D space.
To this end consider an initial scenario of space and time points in an initial system $I^4$ with dimension 4D, and using the coordinates of the metric (\ref{eq1}) for an observed object \textit{G} (Galaxy) at position $r=0$ (indeterminate angles) and time $t=T$, and an observer \textit{O} at position $r=r_1$ (null angles) and time $t=T$ , as show in table 1.
\begin{table}
\tbl{Initial System $I^4$}
{\begin{tabular}{@{}ccccc@{}}
\toprule
System I$^4$ & $r$ & $\theta $ & $\phi $ & $t$ \\ \colrule
G & $0$ & ... & ... & $T$ \\ 
O & $r_1$ & $0$ & $0$ & $T$ \\ \botrule
\end{tabular}}
\end{table}
Then, using the transformations (\ref{eq3})-(\ref{eq5}), (\ref{eq10})-(\ref{eq11}) is obtained the corresponding values in the coordinates of metric (\ref{eq12}) are obtained, in an initial system $I^5\,$ with dimension 5D, as show in the table (2)
\begin{table}
\tbl{Initial System $I^5$}
{\begin{tabular}{@{}cccccc@{}}
\toprule
I$^5$ & $\alpha $ & $\beta $ & $\gamma $ & $\delta $ & $\epsilon $ \\
\colrule
G & $0$ & $0$ & $0$ & $R \cosh \left( \frac T R \right ) (A - B)$ & $R \sinh \left ( \frac T R \right ) (A - B)$ \\ 
O & $0$ & $0$ & $r_1$ & $R \cosh  \left( \frac T R \right ) \left( A - B \sqrt{1 - \frac {r_1^2} {R^2}} \right )$ & $R \sinh \left(  \frac T R \right ) \left( A - B \sqrt{1 - \frac {r_1^2} {R^2}} \right)$ \\
\botrule
\end{tabular}}
\end{table}
that represent the following positions of \textit{G} and \textit{O} in a Cartesian plane for $\varepsilon $ and $\delta $ where
\begin{eqnarray}
p &=& R \cosh \left ( \frac T R \right ) \left ( A - B \right ) \nonumber \\
q &=& R \cosh \left ( \frac T R \right ) \left ( A - B \sqrt{1 - \frac {r_1^2} {R^2}} \right ) \nonumber \\
m &=& R \sinh \left ( \frac T R \right ) (A - B) \nonumber \\
n &=& R \sinh \left ( \frac T R \right ) \left ( A - B \sqrt{1 - \frac {r_1^2} {R^2}} \right ) \nonumber \\
\end{eqnarray}

\begin{figure}[pt]
\begin{center}
\includegraphics[width=0.5\linewidth,clip,trim=240 450 260 330]{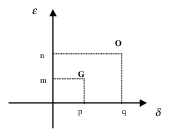}
\end{center}
\caption{Coordinates in $I^4$.}
\end{figure}

We now perform a sequence of transformations, a translation and a rotation which leads to the representation in system $F^5\/$. The translation is defined by the condition that the distance of the origin of the new system $F^5\/$ to $O\/$ is the same as the distance of $G\/$ to the origin of the previous system $I^5\/$.
\begin{figure}[pt]
\begin{center}
\includegraphics[width=0.6\linewidth,clip,trim=240 430 240 330]{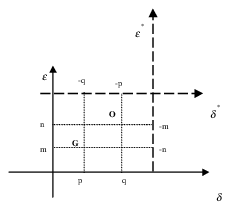}
\end{center}
\caption{Coordinates in $I^4$.}
\end{figure}
This results in the new coordinates
\begin{equation}
\delta ^{*}=\delta -\left( {p+q}\right)  \; , \label{eq13}
\end{equation}
\begin{equation}
\varepsilon ^{*}=\varepsilon -\left( {m+n}\right) \; , \label{eq14}
\end{equation}
and the subsequent rotation is specified by the angle $\sigma=-i\pi $.
\begin{figure}[pt]
\begin{center}
\includegraphics[width=0.7\linewidth,clip,trim=230 420 230 330]{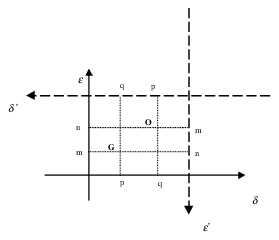}
\end{center}
\caption{Coordinates in $I^4$.}
\end{figure}
The result after two transformations for the two coordinates is
\begin{equation}
\delta^{\prime } = \delta^{*} \cosh (\sigma ) - \varepsilon^{*} \sinh ( \sigma ) = -\delta^{*} \; , \label{eq15}
\end{equation}
\begin{equation}
\varepsilon^{\prime } = \delta^{*} \sinh (\sigma ) + \varepsilon^{*} \cosh (\sigma ) = - \varepsilon^{*} \, . \label{eq16}
\end{equation}

Since the line element (\ref{eq12}) is invariant for Lorentz transformations (\ref{eq15}) and (\ref{eq16}), from (\ref{eq13}) and (\ref{eq14}) one obtains
\begin{equation}
\delta ^{\prime }=-\delta +\left( {p+q}\right) \; , \label{eq17}
\end{equation}
\begin{equation}
\varepsilon ^{\prime }=-\varepsilon +\left( m+n\right) .  \label{eq18}
\end{equation}
Further applying a translation transformation for $\alpha ,\beta $ and $\gamma $, where $\alpha ^{\prime }=\alpha ,\,\beta ^{\prime }=\beta $ and $\gamma ^{\prime }=\gamma -r_1$ one obtains the line element in the new system $F^5\/$
\begin{equation}
ds^2=-d\alpha ^{\prime }{}^2-d\beta ^{\prime }{}^2-d\gamma ^{\prime
}{}^2-d\delta ^{\prime }{}^2+d\varepsilon ^{\prime }{}^2,  \label{eq19}
\end{equation}
where the following table shows the specific coordinates in terms of $A\/$, $B\/$, $R\/$ and $r_1\/$, respectively.
\begin{table}
\tbl{Final System $F^5$}
{\begin{tabular}{@{}cccccc@{}}
\toprule
F$^5$ & $\alpha^\prime $ & $\beta^\prime $ & $\gamma^\prime $ & $\delta^\prime $ & $\epsilon^\prime $ \\
\colrule
G & $0$ & $0$ & $-r_1$ & $R \cosh  \left( \frac T R \right ) \left( A - B \sqrt{1 - \frac {r_1^2} {R^2}} \right )$ & $R \sinh \left(  \frac T R \right ) \left( A - B \sqrt{1 - \frac {r_1^2} {R^2}} \right)$ \\
O & $0$ & $0$ & $0$ & $R \cosh \left( \frac T R \right ) (A - B)$ & $R \sinh \left ( \frac T R \right ) (A - B)$ \\
\botrule
\end{tabular}}
\end{table}
Considering the transformations (\ref{eq3})-(\ref{eq5}), (\ref{eq10})-(\ref{eq11}) in the system $F^5$ one obtains a new system $F^4$ with line element equivalent to (\ref{eq1}) given by
\begin{equation}
ds^2=-\frac{dr\,^{\prime }{}^2}{1-\frac{r\,^{\prime }{}^2}{R^2}}-r\,^{\prime
}{}^2d\theta \,^{\prime }{}^2-r\,^{\prime }{}^2sin^2\theta \,^{\prime
}\,d\phi \,^{\prime }{}^2+\left( {A-B\sqrt{1-\frac{r\,^{\prime }{}^2}{R^2}}}%
\right) ^2dt\,^{\prime }{}^2,  \label{eq20}
\end{equation}
where the spatial and temporal positions of \textit{G} and \textit{O} are given in the following table
\begin{table}
\tbl{Final System $F^4$.}
{\begin{tabular}{@{}ccccc@{}}
\toprule
System F$^4$ & $r^{\prime }$ & $\theta ^{\prime }$ & $\phi ^{\prime }$ & $t^{\prime }$ \\ \colrule
G & $r_1$ & $\pi $ & ... & $T$ \\ 
O & $0$ & $...$ & $...$ & $T$ \\ 
\botrule
\end{tabular}}
\end{table}
Comparing table 1 with table 4 one verifies, that a transformation of coordinates of an initial system $I^4$ with \textit{G} in the origin $r=0$ and \textit{O} in $r=r_1$, to a final system $F^4$ that possesses the same line element (\ref {eq20}) but with \textit{O} in the origin $r^{\prime }=0\/$ and \textit{G} in $r^{\prime }=r_1$, beside the three indeterminate angular values. This
transformation in the context 5D evidences the homogeneity of the 4D interior Schwarzschild solution.

\section{Interpretation}

This result appears paradox in 4D, because it contradicts one of the initial characteristics for the construction of the model that is the variable pressure (no homogeneity). Due to translational symmetry, shown using M-5, one obtains for the pressure \cite{TOL2} of any observer (\textit{G} or \textit{O})
\begin{equation}
p_0=\frac 3{8\pi R^2}\left( {\frac{1-\sqrt{1-\frac{r_1^2}{R^2}}}{3\sqrt{1-%
\frac{r_1^2}{R^2}}-1}}\right) ,  \label{eq21}
\end{equation}
in other words, it only depends on the relative distance $r_1$ between \textit{G} and \textit{O}, but it is constant for both.

Such a result can be interpreted in the context of a theory of the type 5D space-time-matter (STM) \cite{PER,PER2}, since invariance of the metric (\ref{eq12}) should be maintained, and the contradiction is solved admitting the loss of the absolute character of the matter in such theory. In other words, in the theory of the relativity space and time are relative to observers, in a theory STM,  matter is also relative to observers. A theory of this type has as a main characteristics, that it considers matter in a purely geometric fashion, i.e. as a topological quantity. This reasoning is already known in literature, see for instance the theory 5D STM in ref. \cite{WES}. The relative condition of matter in the model can be observed in the
following illustration
\begin{figure}[pt]
\begin{center}
\includegraphics[width=0.4\linewidth,clip,trim=0 740 500 10]{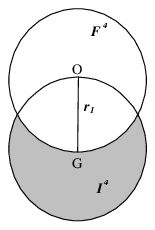}
\end{center}
\caption{Relative condition of the matter.}
\end{figure}
where in the passage of the system $I^4$ (\textit{G }in $r=0$) for the system $F^4$ (\textit{O} in $r^{\prime }=0$), supposing a distribution of radius $r_1$ , the gray part of matter has no influence on metric (\ref{eq20}), but on metric (\ref{eq1}). It is evident in this result the inexistence of an absolute center inside the matter distribution, and consequently \textit{O} and 
\textit{G} lose their uniqueness.

\section{Properties of the Model}

\subsection{Red-Shift}

Another important aspect of the model is the observable red-shift effect that indicates the intrinsic expansion of space (cosmological red-shift), and that from an interior static metric point of view it becomes an effective gravitational red-shift. This result is based on Tolman's work \cite{TOL4} and pointed out by Zeldovich-Novikov \cite{ZEL}, that establishes that external matter of a distribution of spherical symmetry does not possess gravitational effects on the interior matter (considering a radius $r_1$
as separation limit). The observed object \textit{(G)}, is considered with a temporal potential given by a spherical distribution of matter with radius $r=r_1$, i.e. in the surface of the distribution. Therefore we obtain for this potential $g_{tt}=1-\frac{2m}{r_1}\,$. Already the observer \textit{(O)} is considered with a temporal potential given by a spherical
distribution of matter with radius $r_1\longrightarrow 0$, located in $r^{\prime }=0$ , or $g_{tt}\longrightarrow 1$, because it is not
influenced by the external distribution of matter (considering a distribution in a large scale). This can be observed in the following
illustration.
\begin{figure}[pt]
\begin{center}
\includegraphics[width=0.4\linewidth,clip,trim=80 590 300 80]{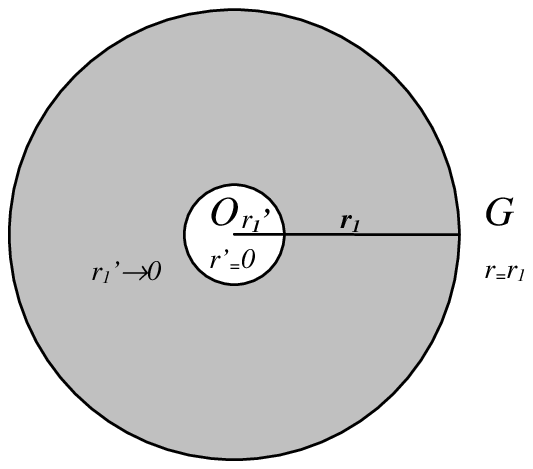}
\end{center}
\caption{Gravitational red shift.}
\end{figure}

Therefore, considering the previous potentials, we obtain to a first order approximation for the red-shift
\begin{equation}
z=\frac{\delta \nu }\nu \sim -\frac m{r_1}  \label{eq22}
\end{equation}
and for (\ref{eq2})
\begin{equation}
z=\frac{\delta \nu }\nu \sim k\,r_1^2\,  \label{eq23}
\end{equation}
which states a law directly proportional to the square of the distance of observation $r_1$, that corresponds to an accelerated red-shift in agreement with astronomical observations \cite{RIE,RIE2,TON}. This result can be experimentally proven by curve adjustment
(regression) coherent with the data for the observed red-shift data and the estimate of matter density that is implicit in the constant $k$. The coefficient $k\,$ of the parabola, determined by $\rho _{00}$ should adjust in a satisfactory way the curve of the observed red-shift. It is worth to notice that the same result can be used in an inverse way to improve the $\rho _{00}$ estimate with bases on the observed red-shift."Double Slit" Young

\subsection{Stability}

The stability of the model can be evaluated by the Buchdahl theorem \cite{BUC} that presents the limit Mass-Radius for the stability of any matter distribution with spherical symmetry and constant density,
\begin{equation}
\frac m{r_1}\le \frac 49\,
\end{equation}
and for (\ref{eq2}) 
\begin{equation}
r_1^2\le \frac 1{3\pi \rho _{00}}\,  \label{eq25}
\end{equation}
as the limit for the radius of the matter distribution (and of observation) as a function of density. This result can be verified experimentally calculating the estimated density in several spheres of astronomical observation (in large scale) of radius $r_1$ (or  consider the mass of all of the objects observed interior to the observation sphere with radius $r_1$). Another experiment would be the impossibility of observing an object at a larger distance than imposed by (\ref{eq25}), taking into account the current estimate of the density of matter $\rho _{00}$ in a large scale. Notice, that the result (\ref{eq25}) can also be used to adjust $\rho _{00}$ if the object of the largest distance $r_1$ still already observed is considered.

\subsection{Static Condition}

An observational result expected for this model, given a static configuration (in large scale) and free from singularities in time
(universe without big bang), one observes matter structures (galaxies) at large distances of conformation, similar to closer galaxies,
suggesting that the creation and destruction of matter, if this mechanism is in action shall be of local nature.

\section{Conclusions}

In the present work we obtained as main results: the possibility of a new static model; the relative condition of matter, or the inexistence of an absolute center in the interior of matter distributions, and the question if in distributions in large scales this
characteristics could affect observation of the local dynamics of observed objects. As seen in Fig. 1, a part of the matter distribution does not influence the metric of the observer; the observation limit given by (\ref{eq25}) can be calculated in units of cosmic time. The accelerated red-shift in spite of the model being static; and some possibilities arise for experimental proofs of the model.
An important aspect of the model is the equivalence of arbitrary matter distribution reference frames,as expected for a model which intend to introduce extensions or generalizations in the framework of the theory of relativity.
Observable quantities or genuine physically relevant variables in such a treatment are, as is the case in the usual 4-dimensional general relativity treatment, quantities which depend typically on the distance of two hyperspace events, in the present work manifest in the equivalent expression for the pressure and density with respect to the initial $I^4$ and the final frame $F^4$.
We have thus shown that there exist distinct but completely equivalent frames where the same physics
may be described. This strongly reminds on a procedure which is today one of the cornerstones of 
quantum field theory and their subsequent developments such as supersymmetric or string theories, i.e. the gauge principle. One may choose a specific gauge without altering the physical content of a model or theory. The same applies here, the connection between $I^4$ and $F^4$ may be considered as a gauge like degree of freedom. If such a reasoning is consitent, then the question arises whether it will be possible to introduce quantization associated to winding, i.e. a loop $I^4 \circlearrowright F^4$, in the same spirit as for instance the Aharonov-Bohm effect may be used to explain quantization of charge.   
Another aspects to be approached in future works are the possible consequent conceptions of modern quantum physics to the proposed extension 5D theory of relativity, such as, Heisenberg's uncertainty principle, discreteness of energy, the wave-particle duality of light and matter, quantum tunneling and interference experiments.

\section*{Acknowledgements}

To professor Bardo Bodmann for constructive collaboration. To UNISC, FACOS, FACCAT, CAPES and IUPAP for financial
support.


\begin{thebibliography}{99}
\bibitem{ABO}  G. Abolghasem, A.A. Coley, and D.J.McManus, 1996, Induced
matter theory and embeddings in Riemann flat spacetimes, \textit{J. Math.
Phys}. 37, 361-373.

\bibitem{BER}  M.Bertola, V.Gorini, U.Moschella, R. Schaeffer,
Correspondence between Minkowski and De Sitter quantum field theory, \textit{%
hep-th}/9906035 v2

\bibitem{BUC}  H. Buchdahl, General relativistic fluid spheres, \textit{%
Phys. Rev}. 116, 1027 (1959).

\bibitem{FRI}  A. Friedmann, \"{U}ber die krummung des raumes, \textit{%
Zeits. f. Physik} 10, pp. 377-386 (1922).

\bibitem{LAC}  M.Lachi\`{e}ze-Rey, The Friedmann - Lemaitr\^{e} models in
perspective, \textit{Astronomy and Astrophysics}, 364, 894-900, 2000

\bibitem{MCM}  DJ McManus, Five-dimensional cosmological models in induced
matter theory, \textit{J.Math. Phys}. 35, 4889 (1994).

\bibitem{PER}  P.H. Pereyra.; B.E.J Bodmann, - Uma transforma\c{c}\~{a}o de
lorentz pentadimensional e suas poss\'{i}veis implica\c{c}\~{o}es- \textit{%
Scientia}, Vol. 12 No. 1- 2001 - Unisinos

\bibitem{PER2}  P.H. Pereyra; B.E.J. Bodmann - Uma extens\~{a}o dimensional
da transforma\c{c}\~{a}o de lorentz e uma aplica\c{c}\~{a}o \`{a}
interfer\^{e}ncia induzida por gravita\c{c}\~{a}o - \textit{Tend\^{e}ncias
em Matem\'{a}tica Aplicada e Computacional} - Vol.3, No.2, (2002), 181-188.

\bibitem{RIE}  A.G. Riess, et.al., Observational evidence from supernovae
for an accelerating universe and a cosmological constant, \textit{Astron. J.}
116 1009 (1998), \textit{astro-ph/}9805201; S. Perl-mutter, et.al.,
Measurements of omega and lambda from 42 high-redshift supernovae,\textit{%
Astrophys. J}. 517 565 (1999), \textit{astro-ph}/9812133.

\bibitem{RIE2}  A.G. Riess, et.al., Type Ia supernova discoveries at z 
$>$ 1 from the Hubble space telescope: Evidence for past
deceleration and constraints on dark energy Evolution,\textit{astro-ph}%
/0402512.

\bibitem{ROB}  H.P.Robertson, On the fundations of relativistic cosmology, 
\textit{Proc.Nat.Acad}. 15, 822 (1929)

\bibitem{SCH}  K. Schwarzschild, \"{U}ber das gravitationsfeld eines
massenpunktes nach der einsteinschen theorie (1916), \textit{arXiv:physics/}%
9912033 v1, 1999 (em ingl\^{e}s)

\bibitem{TOL1}  Tolman, On the estimation of distances in a curved universe
with a non-static line element, \textit{Proc.Nat.Acad}. 16, 511 (1930)

\bibitem{TOL2}  R.C.Tolman, ''Relativity Thermodinamics and Cosmology '', 
\textit{Dover Publications,Inc}.,1987

\bibitem{TOL3}  R.C.Tolman, On the possible line elements for the universe, 
\textit{Proc.Nat.Acad}. 15, 297 (1929)

\bibitem{TOL4}  RC Tolman, Static solutions of Einstein's field equations
for spheres of fluid. \textit{Phys. Rev}., 55:364-373, 1939.

\bibitem{TON}  J.L. Tonry, et.al., Cosmological results from high-z
supernovae , Astrophys. J. 594 1(2003), \textit{astro-ph}/0305008; R.A.
Knop, et.al., New constraints on , and w from an independent set of eleven
high-redshift supernovae observed with HST, \textit{astro-ph}/0309368; B.J.
Barris, et.al., 23 High redshift supernovae from the IfA deep
survey:Doubling the SN sample at z\textquestiondown 0.7, \textit{Astrophys.J}%
. 602 571 (2004), \textit{astro-ph}/0310843.

\bibitem{TRI}  R. Triay, L. Spinelli, R. Lafaye, Framework for cosmography
at high redshift, 1996, \textit{Monthly Not. Roy. Astron. Soc.},279, 564

\bibitem{WES}  P. S. Wesson, ''Space-Time-Matter '', \textit{World
Scientific, Singapore}, 1999

\bibitem{ZEL}  Ya.B.Zel'dovich, I.D.Novikov, ''Stars and Relativity '', 
\textit{Dover P.Inc.}1996
\end{thebibliography}
\end{document}